# Structure Determination in a new Type of Amorphous Molecular Solids with Different Nonlinear Optical Properties: A Comparative Structural Analysis

**J. Link Vasco[1], J. R. Stellhorn[1,3], B. D. Klee[1,4], B. Paulus[1], J. Belz[2], J. Haust[2], S. Hosokawa[5], S. Hayakawa[3], K. Volz[2], I. Rojas León[1,6], J. Christmann[1,6], S. Dehnen[1,6], W.-C. Pilgrim[1,\*]**

[1]Chemistry Department, Philipps-University of Marburg, 35032 Marburg, Germany
[2]Physics Department, Philipps-University of Marburg, 35032 Marburg, Germany
[3]Department of Applied Chemistry, Hiroshima University, Hiroshima, Japan
[4]Institute of Solid-State Physics and Optics, Wigner RCP, 1525 Budapest, Hungary
[5]Department of Physics, Kumamoto University, Kumamoto, Japan
[6]Karlsruhe Inst. of Technology, Karlsruhe, Germany

E-mail: pilgrim@staff.uni-marburg.de

**Abstract**. The microscopic structure of two amorphous materials with extreme nonlinear optical properties has been studied. One of these materials exhibits second harmonic generation, while another material of similar molecular structure emits brilliant white light if being irradiated with a simple IR laser diode. Structural differences were investigated using X-ray scattering and EXAFS combined with molecular RMC. Transmission electron microscopy and scanning precession electron diffraction were used to understand specific structural differences on all length scales, from mesoscopic down to mutual molecular arrangements. Characteristic differences were found at all scales. Close core-core spacing between {SnS} clusters as well as characteristic cluster distortions appear to be characteristic features of the white light emitting material. In the other material, cores are undistorted and core distances are larger. There, the formation of nanocrystalline structures in the amorphous matrix could also be identified as reason for the WLG suppression.

## 1. Introduction

The search for improved materials to generate light has been and is still an active research field. About 25 years ago, these efforts have culminated in the development of the light emitting diode (LED), which has meanwhile become omnipresent everywhere in our daily lives.[1] Typically, LEDs emit a strong line in the dark blue, near UV region resulting from a direct band gap transition. It is then spectroscopically converted to longer wavelengths by dyes and phosphors which cover just the visible range of the electromagnetic spectrum between ~350 to ~800 nm. LEDs are extremely energy efficient and can be tuned to provide any desired color temperature. Another important property of LEDs is that their radiation pattern is almost perfectly Lambertian, i.e., it emits into all directions with virtually the same intensity, which is of great advantage if bright illumination of rooms is desired or if LEDs are used as pixels in flat panel displays where large viewing angles are preferred. For other applications however, a rather point-shaped, laser like radiation characteristics is often wanted. Such light sources also exist and were already developed in the nineteen-seventies as so-called Supercontinuum Emitters (SCEs).[2,3] They are based on strongly non-linear optical (NLO) materials as e.g. YAG, Sapphire, $CaF_2$-crystals, optical fibers or other waveguide-based sources. However, to invoke the NLO effects, high electrical field strengths are needed which are provided by pulsed high-power lasers. These SCEs are therefore heavy and bulky devices with high energy consumption and their use is basically restricted to pure scientific and medical applications.

A few years ago, a group of inorganic-organic hybrid cluster molecules was identified that already exhibit extreme NLO properties when irradiated with just a simple low energy-density continuous wave near infrared (CW-NIR) laser diode.[4,5] These compounds consist of heteroadamantane shaped units of general formula $[(RT)_4E_6]$, where R is an organic ligand, T is a group-14 element, bound to the organic ligand R, and E represents a chalcogen. The heteroatomic composition combined with the wide variability of the organic ligands provides large synthetic variety and meanwhile a huge number of different derivates exist.[6,7] All compounds precipitate as solids, some of which are crystalline, while others show a completely amorphous morphology. Comparing the non-linear optical responses of the different materials, one finds that all crystalline representatives respond as second harmonics generators (SHG) upon IR-irradiation, while most amorphous materials reply as white light generators (WLG). The latter emit warm white light just covering the visible region of the electromagnetic spectrum. Moreover, the emission characteristics of these materials is highly brilliant, retaining the directionality of the driving laser.[4,6] The difference between these two groups is exemplary visualized in Figures 1 (a) and (b) for the two systems $[(PhSi)_4S_6]$ and $[(PhSn)_4S_6]$, in which the molecular units differ only by the exchange of Si by Sn while phenyl groups (Ph) are the organic ligands is in both cases. DFT calculations of these cluster molecules have revealed almost identical molecular structures for both[8] which are displayed in the respective figures. However, the {SiS}cluster solidifies crystalline as is indicated by the corresponding diffractogram in the second row of the Figure, and emits a second harmonic with wavelength 489 nm (2.53 eV) if being irradiated with a CW-NIR line at 979 nm (1.265 eV). On the other hand, the {SnS} cluster precipitates clearly amorphous as can be inferred from the characteristic shape of its X-ray structure factor $S(Q)$ in Figure 1 (b). If irradiated with the same 979 nm laser line it however responds with a brilliant white light emission centered between 400 and 800 nm.

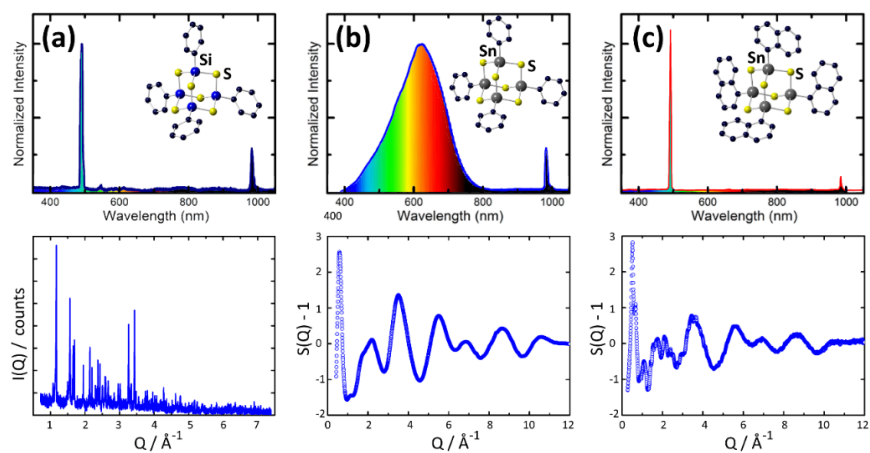

**Figure 1:** NLO-responses from the crystalline material $[(PhSi)_4S_6]$ **(a)** and the amorphous material $[(PhSn)_4S_6]$ **(b)** (top). The driving excitation is visible at 979 nm (1.265 eV) in each spectrum. The 2$^{nd}$-harmonics of **(a)** is clearly seen at 489.5 nm (2.53 eV), while **(b)** depicts a broad white spectrum. The respective X-ray patterns are also shown below indicating that the SHG-material is clearly crystalline while the WLG material shows the typical structure factor $S(Q)$ of an amorphous solid. **(c)** shows the NLO-response from $[(NpSn)_4S_6]$ indicating SHG, although the X-ray structure factor clearly designates an amorphous solid.

Yet, the underlying process for white light generation is still unclear, as is the reason why some cluster systems crystallize while others exclusively solidify amorphous. However, the observation that WLG is never observed in crystalline materials indicates that the effect must be related to specific structural correlations or degrees of freedom that are only attainable in a sufficiently disordered state. This raises the question of how the disordered state is characterized in these systems. This question can only be understood if both the mutual arrangement of the molecules on a microscopic scale in the sub-nano range is known, as well as morphological variations on mesoscopic lengths. This is the only way to understand the relationships between order and disorder in these materials, which is essential for an understanding of the structure-property relationships in view of the nonlinear optical behavior. Among all the materials synthesized so far, there are very few which appear to be amorphous but nevertheless react as SHGs upon irradiation, as is shown in Fig. 1 (c) for the [(NpSn)$_4$S$_6$]-cluster with naphtyl (Np) ligands as the organic component. Its X-ray scattering pattern clearly identifies it as a non-crystalline material, which is evident from the shape of the measured structure factor $S(Q)$ in the figure. The fact that some few SHG materials are amorphous opens up the possibility of detecting precisely the microscopic structural differences without them being masked by the different morphology. We therefore performed structural studies on the amorphous SHG [(NpSn)$_4$S$_6$] and the WLG [(PhSn)$_4$S$_6$] to search for microscopic spatial differences in order to identify the specific structural features of the WLG materials. For this, we performed measurements of the EXAFS (Extended X-ray Absorption Fine Structure) function $\chi(k)$ and the static X-ray structure factor $S(Q)$, and analyzed the resulting data by means of molecular Reverse Monte Carlo simulations (m-RMC). Scanning transmission electron microscopy ((S)TEM) measurements, combined with scanning precession electron diffraction ((S)PED), were also performed to unravel the relationships between morphology and microscopic structure to understand why some materials do not emit white light despite their amorphous appearance. Here, we give a comprehensive overview of the results from the selected systems [(PhSn)$_4$S$_6$] and [(NpSn)$_4$S$_6$], which are shown in Figs. 1 (b) and (c).

## 2. Methods

*2.1. Sample preparation*
The molecular {SnS}-cluster materials [(PhSn)$_4$S$_6$] and [(NpSn)$_4$S$_6$] were prepared according to literature,[9,10] by reacting organotin trichlorides (RSnCl$_3$) with sodium sulphides where R was either Ph (-C$_6$H$_5$) or Np (-C$_{10}$H$_7$). All synthesis steps were performed under Argon atmosphere and the substances were obtained as amorphous white stable and non-hygroscopic powders. Final product analysis was performed by NMR and mass spectrometry, and preliminary morphology studies performed by X-ray diffractometry to clarify their crystalline or amorphous nature. The molecular structures of the substances were further elucidated by density functional theory (DFT) calculations[8] supporting an inversion-free heteroadamantane-type molecular structure with (idealized) T$_d$ symmetry.

The density of the solid samples was measured with an AccuPyc II 1340 Gas Displacement Pycnometry System (micromeritics) using Helium gas. The measurements consisted of 30 purge steps, following 50 measurements per sample which were averaged to give a density accuracy of up to two decimal places. For [(PhSn)$_4$S$_6$] a value of 2.00 g/cm$^3$ was found, while the density of [(NpSn)$_4$S$_6$] was determined to 1.79 g/cm$^3$.

*2.2. X-ray scattering*
High precision X-ray scattering data of the [(PhSn)$_4$S$_6$] material were measured in transmission geometry at beamline P02.1[11] of the PETRA III synchrotron at DESY, Hamburg using a primary energy of 59.87 keV. Scattered X-rays were collected using a two-dimensional position sensitive detector with 2048×2048 pixels of size 200×200 μm$^2$. The DAWN software package was employed to convert the 2D image into scattering pattern.[12] Distance between sample and detector was set to 240.2 mm and the sample was positioned in front of a detector corner to obtain maximum angular range. The [(NpSn)$_4$S$_6$] sample was explored on an inhouse Bruker D5000 diffractometer equipped with a Goebel mirror to optimize the primary beam (Mo $K_\alpha$, 17.44 keV). Both samples were confined in boro silicate X-ray capillaries of 1 mm outer diameter and 0,01 mm wall thickness. Measured scattering intensities were corrected for background- and air-scattering, self-absorption, polarization and Compton contribution and then normalized to $S(Q)$.

*2.3. EXAFS experiments*
Tin-$K$ edge EXAFS (29.2 keV) for both samples were obtained at beamline P65,[13] also located at PETRA III, while sulfur-$K$ edge EXAFS (2.47 keV) were measured at beamline BL-11[14] at the HiSOR facility of the Hiroshima Synchrotron Radiation Center in Japan, which is designed to maximize the beam intensity on the sample within the 2–5 keV region. At BL-11, the measurements were carried out at room temperature and the sample was directly measured, sandwiched between two sulfur free polypropylene foils. At P65 the samples were mixed with graphite and pressed to pellets. All scans were performed in transmission mode. The absorption spectra were normalized and background was calculated using the AUTOBK algorithm. The data were finally analyzed using the Demeter software package (Athena and Artemis).[15]

*2.4. Molecular Reverse Monte Carlo simulations*
An existing RMC-code from the RMC POT++ program package,[16] which already provides the ability to group atoms as rigid molecules and move them along molecular translational and rotational degrees of freedom, was accordingly manipulated for our needs. For the two materials, no crystal structures were known that could have been used as starting configurations for the simulations. Therefore, random molecular arrangements had to be generated which, on the one hand, had to correspond to the real particle densities and, on the other hand, should no longer contain any overlapping molecules. In the original script of the RMC_POT++ package, all atom pairs that violated these cut-off conditions were identified, and all molecular moves that resulted in such pairs were prohibited. However, random initial configurations of large molecules inevitably contain atomic overlaps, and any attempts to move entire molecules unavoidably lead to new overlaps. It is therefore impossible to disentangle a random initial configuration of larger molecules with such strict constraints. We therefore used a less stringent procedure in which a quantity S was defined as a measure for cut-off violations which had to be minimized. However, moves resulting in overlaps were accepted as long as no additional overlaps were created, i.e. all molecular motions were allowed as long as the value of S was not increased. The algorithm was further modified in that complex histogram calculations were restructured and parallelized and a subdivision of the simulation box was introduced enabling the control of cut-off conditions to be limited only to the immediate vicinity of a moving molecule. Finally, these improvements led to an increase in computational speed by almost a factor of 400.[17]

In each simulation, 216 copies of the DFT calculated {SnS}-clusters were moved under periodic boundary conditions in cubic simulation boxes whose sizes were chosen as to match the respective sample densities. The experimental X-ray $S(Q)$s and EXAFS-$\chi(k)$s were used as experimental boundary conditions to which the corresponding functions calculated from the simulated structures should ideally converge. First, the rigid molecules calculated with DFT[8] were moved along their translational and rotational degrees of freedom until best possible agreement between simulated and measured data was achieved. Then, after this rigid m-RMC simulation, a second dynamic simulation was performed where sulfur and tin atoms were additionally allowed to slightly vary their coordinates inside the cluster-cores within certain limits.

*2.5. Transmission Electron Microscope and electron scattering studies*
The advantage of (S)TEM is that possible mesoscopic crystalline inclusions in an amorphous matrix can be identified locally, which is not possible with scattering methods that can only provide the ensemble average of an irradiated sample. On the other hand, (S)PED allows to selectively target such areas and collect meaningful information about them using local electron diffraction. Thus, as in conventional diffraction experiments, structural properties can be obtained on a molecular basis from sample regions only nanometers apart. The spatial resolution over the sample volume is thus much higher than in conventional scattering experiments with neutrons or X-rays.

The TEM measurements were performed using a conventional JEOL JEM-3010 at 300 kV equipped with a TVIPS X416F-ES camera providing single electron sensitivity. Due to this camera it could be operated under low-dose conditions. For the measurements and location images these were on the order of $10^{-4}$ e/Å$^2$ for the low magnification mode and ~$10^{-2}$ e/Å$^2$ for higher magnifications. Measurements were performed at room temperature to obtain comparable data to the above-mentioned X-ray diffraction studies.

(S)PED measurements were performed using the NanoMEGAS P2010 beam scanning/precession system installed on a double aberration corrected JEOL JEM-2200FS. This system produces a focused convergent probe with variable precession angle. An angle of approximately 1.0 degree was used. The probe was scanned over the sample, and the diffraction planes were recorded for each scan point by a camera pointed at the microscope's built-in phosphor screen. Thus, the pixel information corresponds to the slightly convergent (about 0.8 mrad) diffraction pattern at a camera length of about 53 cm and originates from the area under investigation. The spatial resolution is determined either by the resolution of the scan point or by the physical size of the probe, which for this experiment was measured to be about 1.8 nm. Further details on these experiments are given elsewhere.[18, 19]

**Results and Discussion**

Fig. 2 shows experimental EXAFS results (symbols) for the amorphous SHG- and WLG -materials [(NpSn)$_4$S$_6$] and [(PhSn)$_4$S$_6$], obtained at the Sn $K$- and the S $K$-edge. All spectra can be fitted reasonably well by the EXAFS function $\chi(k)$. S-Sn and S-S scattering paths were used for the S $K$-edge data and Sn-S, Sn-Sn and Sn-C paths at the Sn $K$-edge. Mean displacements were also used as fitting parameters. Fit windows were defined between 1.2-4.0 Å for the Sn $K$-data, and between 1.5-3.8 Å for the S $K$-data. Structural parameters from the DFT-calculated[8] molecules shown in the figures, were used as starting parameters for the fits. The obtained results are displayed as red lines. They are close to the curves expected from the single molecule DFT-calculations. However, the fits at the sulfur edges are considerably better for the SHG material with organic naphtyl ligands than for the phenyl containing WLG material. This is also apparent from the goodness of fit values shown in the figures as red numbers. Their difference indicates that the calculated molecular structure of the WLG-cluster seems to experience stronger modifications if being transferred into the dense amorphous phase than the SHG-cluster. Since this difference is only visible in the sulfur EXAFS, it is reasonable to assume that the structural effect is exclusively related to the sulfur-sulfur correlations. Therefore, an additional S-S fitting path was introduced into the fitting procedure for [(PhSn)$_4$S$_6$], which indeed resulted in a significant improvement of the fit quality. This is represented in Fig. 2 by the blue line in the spectrum for the phenyl cluster and also reflected by the improved goodness of fit value (blue number in the figure) which is now of the same order of magnitude as for the SHG-material. The additional fitting path indicates that an additional sulfur atom is situated nearby either due to a distortion of the molecular structure or due to an additional intermolecular correlation. The additional scattering path yields, a S⋯S spacing of 3.6 Å. It must however be stated that the inclusion of an additional fitting path causes an increased dependence among the fitting parameters. Therefore, additional information was needed to confirm the reliability of this procedure.

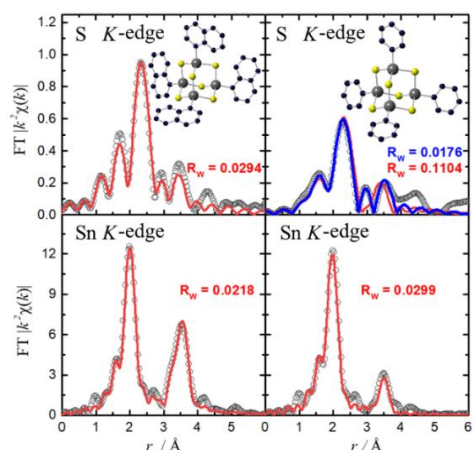

**Figure 2:** Experimental real space EXAFS data (symbols) obtained at the sulfur $K$-edge (top) and the tin $K$-edge (bottom) of the [(NpSn)$_4$S$_6$] (left) and the [(PhSn)$_4$S$_6$] (right), respectively. Solid red lines represent fits to the data using the respective DFT-calculated models as shown in the Figures. Blue line in the spectrum for the phenyl cluster denotes an extended fit using an additional S-S scattering path to fit the data.

Fig. 3 shows the results of the m-RMC simulations for the amorphous [(PhSn)$_4$S$_6$] and [(NpSn)$_4$S$_6$] materials, using the X-ray scattering data already displayed in Figure 1 and the EXAFS data from the Sn $K$-edges as constraints for the simulation. The thinner blue line in the results for the phenyl cluster represents a simulation attempt using rigid molecular clusters, based on the DFT-calculated structural model.[8] It can be seen that the essential features of $S(Q)$ are reasonably well reproduced, but the agreement with the data is only moderate. E.g., a phase shift can be observed for $Q$-values above about 8 Å$^{-1}$ indicating discrepancies between the DFT-calculated structure model and the real shape in the amorphous matrix which complies with the above interpretations from EXAFS. The observed difference is even more pronounced in the Sn $K$-edge EXAFS data, where the corresponding blue curve deviates considerably from the experimental findings (symbols).

Another simulation was hence performed where sulfur and tin atoms could vary their coordinates in the cluster cores within given limits: The Sn-S bond was allowed to vary more or less freely between 2.05 and 2.65 Å, since its contribution to the X-ray and EXAFS data is large, yielding sufficient information density for a reliable simulation. The Sn–C bond was stronger constrained to 2.05 - 2.25 Å due to its smaller weighting and thus the smaller experimentally information density. The C atoms were not allowed to move intramolecularly, except that the organic groups were allowed to rotate around the Sn-C bonds. The Sn atoms thus always remained close to their original coordinates, which ensured an intact overall molecular structure during the simulation process. The results of this simulation are shown in Fig. 3 by the full red lines. Both, $S(Q)$ as well as $\chi(k)$ calculated from the atomic coordinates are now in nearly quantitative agreement with experiment. This latter simulation procedure was then applied to the [(NpSn)$_4$S$_6$] system, yielding similar good results for $S(Q)$ and $\chi(k)$ as is also shown in Fig. 3.

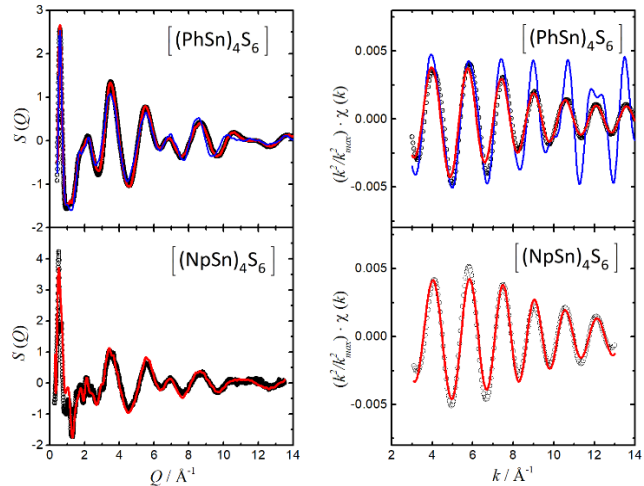

**Figure 3:** Comparison between experimental and m-RMC simulated $S(Q)$ and Sn $K$-EXAFS functions for the WLG [(PhSn)$_4$S$_6$] and the SHG [(NpSn)$_4$S$_6$]. Symbols are experimental data. The blue lines in the graphs for the phenyl-cluster represent simulation results using the rigid DFT-calculated molecule.[8] Red lines in all graphs are results of the dynamic simulation, where Sn and S atoms were allowed to slightly vary their positions inside the cluster cores.

The influence of the dynamic m-RMC on the structure of the cluster cores is illustrated in Figs. 4 by the intramolecular partial pair distribution functions (PPDF) as obtained from the simulation boxes. The blue dashed vertical lines respectively indicate the Sn⋯Sn, S⋯S and Sn⋯S spacings expected from the undistorted DFT-calculated clusters.[8] The curves represent the results from the dynamic m-RMC simulations. The red solid lines, belonging to the right-hand scales are the so-called running coordination numbers defined as the integral from 0 up to a given value of $r$ over the respective radial distribution functions (RDF), $4\pi \cdot n_k \cdot g(r) \cdot r^2$, with $n_k$ being the particle density of element k. It determines the number of neighboring atoms hidden under a PPDF peak. It can be seen from Fig. 4 (a) that the {Sn$_4$S$_6$} cores of the amorphous [(PhSn)$_4$S$_6$] material are considerably distorted. The Sn⋯Sn correlation peak in $g_{Sn\text{-}Sn}(r)$, is asymmetrically broadened, revealing a deformation of the originally tetrahedral Sn$_4$ frame. The first peak in $g_{Sn-S}(r)$ (Fig. 4 (a), center) represents the three sulfur atoms at 2.44 Å to which each Sn atom is chemically bonded. In the undistorted model cluster, three further S atoms are situated at 4.63 Å as second next neighbors (blue dashed vertical line). However, in the amorphous solid this correlation is split into three components between 3.5 and 5.5 Å, containing these three neighbors. The structure model also suggests four next neighbors in $g_{S-S}(r)$ at 4.01 Å (1$^{st}$ vertical dashed line in $g_{S-S}(r)$) which is also split into two distances below and above this value. The running coordination number exactly reveals that two of the four S-neighbors are shifted to smaller distances (3.68 Å) over a narrow correlation range, while the two other atoms are situated farther away (4.18 Å), distributed over a wider $r$-range. The second S⋯S spacing originally located at 5.7 Å in the rigid DFT-calculated cluster is considerably broadened and shifted to smaller distances (5.51 Å).

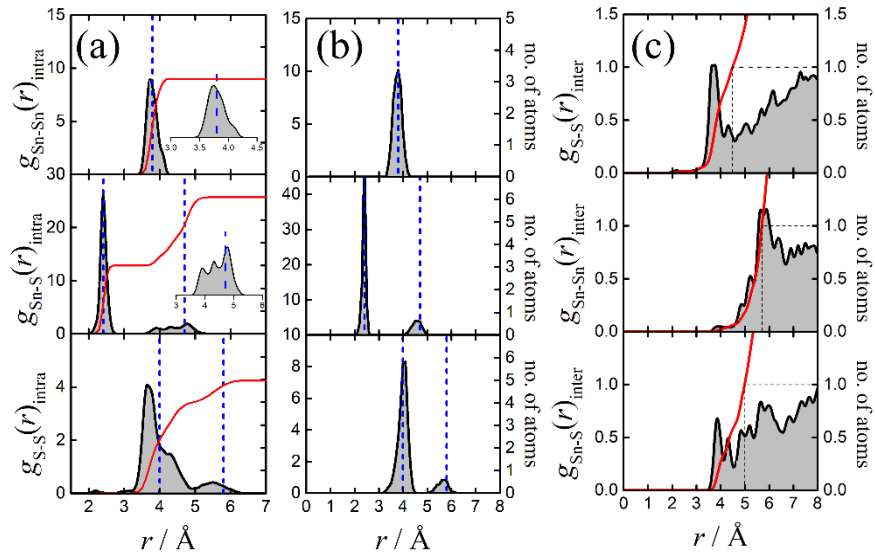

**Figure 4:** Comparison between the intramolecular partial correlations functions for the WLG [(PhSn)$_4$S$_6$] **(a)** and the SHG [(NpSn)$_4$S$_6$] **(b)**. The insets in the two upper diagrams in **(a)** show enlargements of the respective peaks in the graphs. The blue dashed lines in the graphs represent internal distances in the cluster cores as expected from the DFT calculated structures.[8] **(c)** shows the intermolecular partial correlation functions for the [(PhSn)$_4$S$_6$] system.

The PPDFs of the amorphous SHG-system [(NpSn)$_4$S$_6$] as obtained from the m-RMC are displayed in Fig. 4 (b). Here however, no molecular distortions are observed. All correlation peaks are close to the values predicted by the DFT-calculated model and no splitting or asymmetric broadening of the correlation peaks is found. Hence, the cluster cores of this SHG-material are largely undistorted and closely resemble the DFT-calculated model,[8] which is consistent with the above EXAFS analysis.

The mutual spatial arrangement of phenyl- and the naphtyl-clusters in the simulation boxes are represented in the upper insets of Figs. 5 (a) and (b), where the positions of the molecular centers of mass are displayed, respectively. The graphs below are the pair distribution functions (PDF), $g_m(r)$, of these centers. The [(PhSn)$_4$S$_6$] PDF indicates that molecular centers do not approach closer than 6 Å. Above this value a steep correlation rise occurs forming a pronounced peak centered between 6 and ~9 Å. Further increased correlation exists between ~11 and ~14 Å.

For distances between 6 and 7 Å dimeric structures are exclusively found in the simulation box. They represent about 20 % of all molecules. They are indicated by the red bonds in Fig. 5 (a). A typical dimer from the RMC ensemble is shown in Fig. 6 (a). The mutual molecular arrangement is an alternating staggered configuration where the ligands of one molecule are located in the voids between those of the other molecule allowing maximal approximation of the cores. A preference of this conformation for {SnS} and {SiS} clusters with phenyl ligands was found in DFT based binding energy calculations, where cluster dimers were studied as minimal models of the amorphous state,[6,20] and where intra dimer distances between 6.0 and 6.5 Å were proposed for [(PhSn)$_4$S$_6$]. Here, we find an average dimer spacing of 6.75 Å, which is slightly larger, and can be attributed to the fact that the dimer interaction in a real solid is also shared with other molecules. A staggered alternating arrangement between dimers is also found in the crystal structure of [(PhSi)$_4$S$_6$].[8]

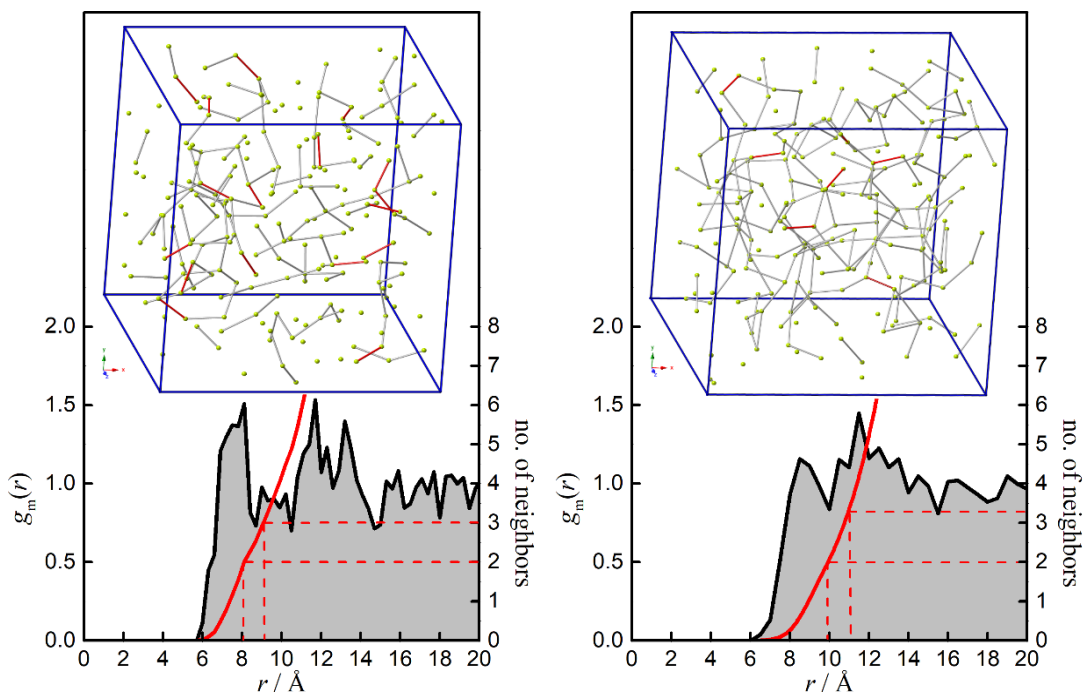

**Figure 5:** $g_m(r)$, of the molecular centers as obtained from m-RMC simulations on [(PhSn)$_4$S$_6$] **(a)** and [(NpSn)$_4$S$_6$] **(b)**. Red full lines represent the integrals over the RDFs ($4\pi n r^2 g(r)$) to give average numbers of surrounding molecules. Inset shows distribution of the molecular centers in the simulation box. In **(a)** dimer bonds are drawn in red for $r$ values up to 7 Å. Other bonds are drawn up to 8.5 Å. In **(b)** the red dimer bonds are drawn for spacings up to 8 Å. Grey bonds show correlations up to 11 Å.

The RDF-integral (red lines, right scales) shows that each molecule is on average surrounded by two neighbors at 8.0 Å, indicating that the first maximum in $g_m(r)$ may mainly result from chainlike structures. Indeed, such structures dominate the mutual alignments in the simulation box up to this correlation length. A linear tetramer chosen from the RMC ensemble is exemplary shown in Fig. 6 (b). The respective intermolecular distances are given by red numbers. Again, we find alternating staggered mutual alignments of the organic ligands. At larger distances, up to 9 Å the running coordination number indicates three neighbors on average. This value lies between the first two maxima in $g_m(r)$. Here, the chains begin to branch as is indicated by the grey bonds between the molecular centers in the simulations box of Fig. 5 (a). From Fig. 6 (a) and (b), the distortion of the cluster cores is already visible, which obviously results from modified sulfur positions. Some inter- and intramolecular sulfur spacings are given in the figure, indicating that the overall difference between intra- and intermolecular S–S spacings seems to vanish. It appears as if the sulfur atoms tend to distribute themselves uniformly. This is also apparent from the intermolecular PPDFs shown in Fig. 4 (c). A pronounced correlation peak centered at 3.7 Å is clearly visible in $g_{S-S}(r)_{inter}$. The integral over the RDF indicates that every cluster core of the RMC ensemble is on average surrounded by one S atom from another cluster core between 3 and 4.18 Å, 75% of which are situated in the segment under the peak at 3.7 Å, which is in good agreement with the spacing found for the additional scattering path in the sulfur $K$-edge EXAFS analysis alone. This value is well inside the range of the intramolecular S···S spacings. A similar strong correlation is found in $g_{Sn-Sn}(r)_{inter}$ which however extends to considerably larger values, and involves only about half an atom on a comparable length scale. Much weaker intermolecular correlations exist between Sn and S atoms.

The mutual spatial situation of the centers of mass for the SHG material [(NpSn)$_4$S$_6$], as obtained from the dynamic m-RMC simulation, is displayed in Figure 5 (b). Here, a first maximum in $g_m(r)$ is centered around 8.6 Å, which is at higher distance as for the WLG-material, indicating that cluster cores are farther apart in the SHG case. Also, the increase in correlation above 6 Å seems to be shallower than in the WLG case, suggesting that repulsive forces are weaker and nearest neighbor distances are spread over wider ranges. A second maximum is located nearby between about 11 and 14 Å. Only 6.5% of the molecules form dimer pairs in a correlation range between 6.5 and 7.5 Å (red bonds in Fig. 5 (b)), which is considerably less than for the WLG material. At higher correlation length the number of neighbors increases rapidly and at ~10 Å, the integral over the RDF indicates two next neighbors on average. In fact, longer chains are found in the simulations box up to this length scale, which are however already strongly branched, and at 11 Å the running coordination number indicates already more than three next neighbors. Here, in contrast to the WLG material, such values lie in the range of a distinct broad peak in $g_m(r)$ indicating that interconnection between the molecular centers on this length scale has now formed a dense network as is shown by the grey bonds displayed in the simulation box of Fig. 5 (b). Fig. 6 (c) shows a typical example for the mutual arrangement of five [(NpSn)$_4$S$_6$] molecules arbitrarily selected from the m-RMC simulation box. The central molecule is shaded blue for better distinction. It is surrounded by two other molecules about 8.5 Å apart and by two further molecules at a about 11 Å.

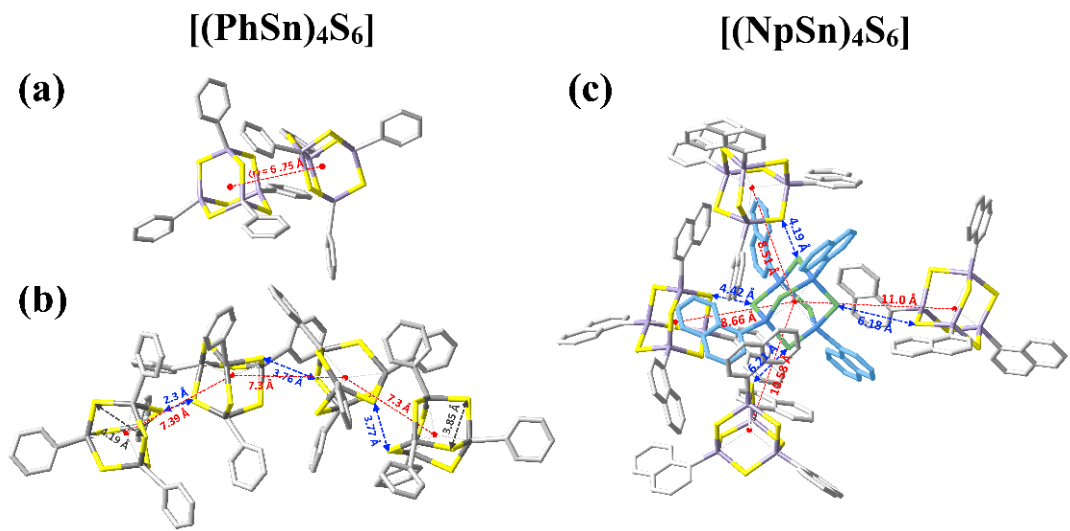

**Figure 6:** Arbitrarily chosen mutual molecular arrangements taken from the m-RMC simulation boxes. Yellow: positions of S atoms, purple: positions of Sn atoms, grey: organic ligands. **(a)** [(PhSn)$_4$S$_6$]-dimer from the m-RMC simulation box. Average spacing between molecular centers is displayed by red numbers. **(b)** [(PhSn)$_4$S$_6$]-tetramer from the m-RMC simulation box. Red numbers indicate spacings between molecular centers, blue numbers and arrows are intermolecular S-S distances. Some intramolecular S-S spacings are given in black. **(c)** Molecular crosslinking between five [(NpSn)$_4$S$_6$] molecules. The central molecule (shaded blue for better distinction) is surrounded by two other molecules about 8.5 Å apart and by two more at a about 11 Å. Blue and red numbers denote spacings between intermolecular centers and S-S atoms, respectively.

Both, EXAFS analysis and m-RMC simulations indicate that the adamantane-like molecular cores in the WLG material are distorted, while in the SHG material molecules are undistorted. To reproduce the X-ray and EXAFS patterns, more of the shorter S-S distances are required for the WLG than can be provided by the undistorted adamantane cluster. Therefore, in the simulation, the sulfur atoms move out of their original positions to form shorter intermolecular S-S distances. This is indicated by the intense intermolecular S-S correlation peak in Fig. 4 (c). In the real amorphous WLG material, the sulfur atoms seem to strive for a uniformly distributed sulfur network, but since the atoms are tightly bound in their molecular framework, this results in a distortion of the molecular cores. Since no chemical bonds exist between the sulfur atoms, it is tempting to identify the formation of such a sulfur mesh as a vibrational network. Such a network could be the source for an enhanced density of vibrational states with a broad range of *k*-values which could explain the observed high receptivity of the WLG materials for infrared radiation.

All cluster nuclei are distorted differently leading to strong non-uniform spatial fluctuations in the interaction forces that suppress crystallization. Strong isotropic core-core interactions were previously already suspected to hinder crystal formation.[20]

Our results and interpretations presented so far can however not answer the important question, why the [(NpSn)$_4$S$_6$] system does not act as a WLG although it appears to be amorphous. Also, the scattering law of [(NpSn)$_4$S$_6$] does not show any distinct Bragg peaks and resembles the typical $S(Q)$ of a disordered condensed phase. Nevertheless, it should be noted that it also exhibits peculiar fluctuations between one and three Å$^{-1}$ that are untypical for fully disordered systems like liquids and glasses, where $S(Q)$ is a rather smooth function. Therefore, a comprehensive study was also carried out to explore the structural properties covering the range from mesoscopic down to microscopic scales using (S)TEM, combined with (S)PED.[18] The latter allows to perform electron diffraction experiments at different sample positions with a spatial resolution down to about 1.5 nm.

Figure 7 (a) shows a high angle annular dark field overview image from a [(NpSn)$_4$S$_6$] sample obtained by (S)TEM. The tin-sulfide compound is displayed by the bright areas in the image. Two different modifications of the compound can be identified: large, µm-sized round particles and also significantly smaller, rod-like units. Fig. 7 (b) shows the virtual bright field image reconstructed from the diffraction patterns intensities obtained across the scanned area. The dashed box indicates the final (S)PED acquisition data set region that corresponds to the data shown in Figures 8 (d–f).

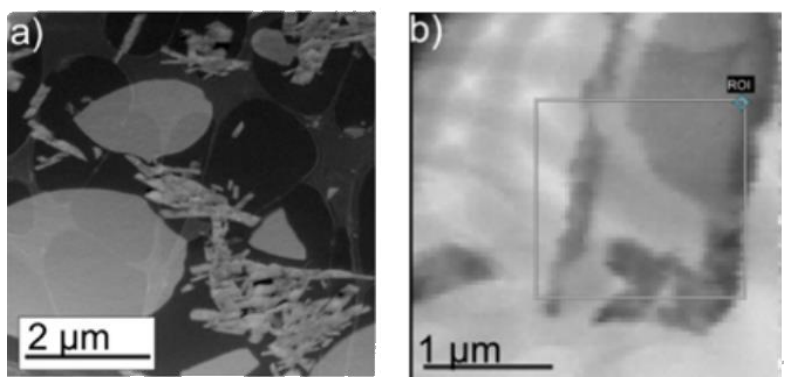

**Figure 7:** (a) High angle annular dark field (HAADF) (S)TEM images of differently sized [(NpSn)$_4$S$_6$] particles (bright regions) embedded in an epoxy matrix (black) on lacey carbon support (dark grey). (b) Virtual bright field image generated from the (S)PED dataset showing an inverted contrast compared to the HAADF. {SnS}cluster region is darker than the background. From this image a small subset was generated for the indicated region of interest (ROI). Diffraction patterns from different regions of that data set are shown in Fig. 8.[18]

Diffraction patterns (DP) were recorded and stored for all scan points in this region. From selected pixels of the DPs for each scan point (see Figure 8 (a–c)) virtual dark field (VDF) images were generated which are proportional to their intensity (Figure 8(d–f)). Most of the scan points (a, d) confirm the amorphous structure, also inferred from the X-ray experiments. However, (S)PED clearly reveals nanoscale regions exhibiting distinct diffraction spots (Fig. 8 (b-c) and (e–f)). These crystallites, indicated by cyan and orange arrows, are about 50 to 150 nm in size, and are mostly found around the rodlike particles (cyan) as well as on the edges of the round particles (orange). The (S)PED analysis of the WLG [(PhSn)$_4$S$_6$] does not show any crystalline inclusions and a fully amorphous morphology is confirmed. Electron scattering pattern of the amorphous areas from low dose TEM measurements could be reduced down to the absolute $S(Q)$ level from which PDFs were also obtained by Fourier transform.[18,19] Within the limits of the slightly different relative scattering lengths, they are in good agreement with the findings from direct X-ray scattering.[21]

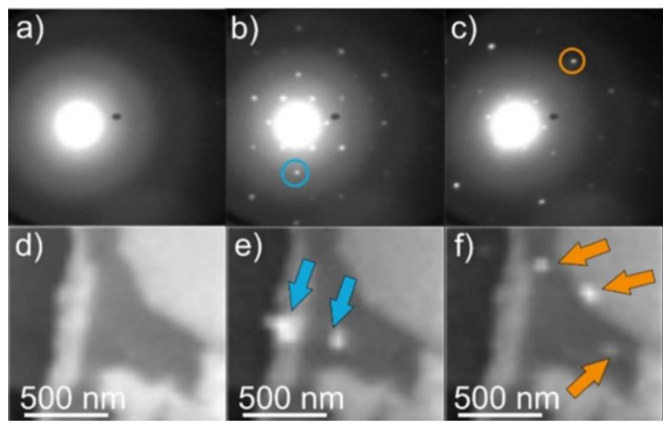

**Figure 8:** The diffraction pattern in (a) originates from amorphous regions of the [(NpSn)$_4$S$_6$] material whereas (b) and (c) originate from regions highlighted in (e) and (f), respectively. By selecting specific regions in the diffraction pattern of the data set, virtual dark field images can be generated. An arbitrary pixel taken from (a) shows a dark field map of the scanned area, as depicted in (d). In contrast, a region chosen on a diffraction spot like indicated in (b) and (c) highlights regions that generate these diffraction spots. It is apparent that the origin of the diffraction spots stems from small crystalline regions.[18]

Apparently, on mesoscopic length scales, [(NpSn)$_4$S$_6$] consists of crystalline and amorphous regions but the amorphous regions predominate by far. The tendency for crystallization may stem from the fact that the molecular cluster cores are undistorted. Thus, identical directed interactions exist between all clusters. On the other hand, the crystallization is also sterically hindered by the bulky organic ligands, which are free to rotate about the Sn-C axis. As a result, the cluster cores cannot approach close enough for effective crystallization. Also, nanometer-sized crystalline spots are found to exist in the amorphous matrix. The unusual intensity oscillations observed in $S(Q)$ of [(NpSn)$_4$S$_6$] between 0.1 and ~3 Å$^{-1}$ could be remnants of Bragg peaks originating from such spots. Due to the extremely small crystallite sizes in the nm range, such peaks were extremely Scherrer broadened and can therefore not be resolved in a conventional scattering experiment.

**Conclusions**

The structural properties of the WLG material [(PhSn)$_4$S$_6$] and the SHG material [(NpSn)$_4$S$_6$] were investigated in an extensive structural study, addressing correlations from the micrometer range down to inter- and intramolecular scales. Clear structural differences exist between the two materials on all scales, which on one hand justifies why they don't condense crystalline, and on the other hand also provides indications for their different optical behavior. On molecular scales, EXAFS and X-ray scattering reveal pronounced molecular distortions for the WLG material, which can be attributed to variations of the sulfur positions in the cluster core. Inter- and intramolecular sulfur distances are similar, suggesting a sulfur network. Since there are no chemical bonds between these atoms, a pure vibrational network may be speculated, which could contribute to an increased vibrational density of states which could explain the high IR receptivity of the WLG.

No molecular distortions are observed in the SHG material. Here, larger distances between the {SnS} cluster nuclei are found as a result of steric hindrance by the more voluminous organic naphthyl ligands. These dominate the intermolecular interaction, but also suppress crystallization. While (S)TEM studies of the [(PhSn)$_4$S$_6$] material show a homogeneous amorphous matrix on the micrometer scale, different morphologies are found for the [(NpSn)$_4$S$_6$] on this length scale: larger round and rod-shaped particles can be distinguished, respectively. Electron diffraction with spatial resolution in the nanometer range on these different domains demonstrate the existence of nanocrystalline domains in the otherwise amorphous matrix, suggesting that the crystallization suppression by the organic ligands is weaker in this system than in the [(PhSn)$_4$S$_6$] material, where the strong distortion of the cluster cores was made responsible.

**Acknowledgments**

We acknowledge funding by the German Research Foundation (Deutsche Forschungsgemeinschaft, DFG), Grant No. 398143140, related to the Research Unit FOR 2824. The authors also acknowledge the great working conditions and support of the following large-scale facilities: German Electron Synchrotron (Deutsche Elektronen Synchrotron, DESY, a member of the Helmholtz Association HGF), beamlines P65 (proposal ID I-20190122), P02.1 (proposal ID RAt-20010143), and the HiSOR facility of the Hiroshima Synchrotron Radiation Center (BL-11, proposal No. 20AG034).